\documentclass[12pt]{article}

\usepackage{mathptmx}      
\usepackage{amsmath}
\usepackage{amssymb}
\usepackage{graphicx}
\usepackage{natbib}
\usepackage{color}
\usepackage{verbatim}
\usepackage{url}

\newcommand{\A}{p_a}
\newcommand{\B}{p_b}
\newcommand{\C}{p_c}
\newcommand{\D}{p_d}

\newcommand{\e}{e}
\newcommand{\s}{t}
\newcommand{\g}{g}
\newcommand{\F}{\mathcal{G}}  
\newcommand{\G}{G}            
\newcommand{\w}{\omega}       
\newcommand{\h}{h}            
\newcommand{\var}{\sigma}
\newcommand{\pd}{\partial}
\newcommand{\R}{L}
\newcommand{\PP}{P}
\newcommand{\I}{\alpha}
\newcommand{\p}{p}
\newcommand{\FSLM}{FSLM}
\newcommand{\FFCM}{FFCM}
\newcommand{\pr}{\mathbf{P}}

\begin{document}
\title{Length distribution of sequencing by synthesis: 
fixed flow cycle model
}


\author{Yong Kong\\
Department of Molecular
Biophysics and Biochemistry\\
W.M. Keck Foundation Biotechnology Resource Laboratory \\
Yale University\\
333 Cedar Street, New Haven, CT 06510\\
email: \texttt{yong.kong@yale.edu} }

%

\date{}

\maketitle


\begin{abstract}
Sequencing by synthesis is the underlying technology for many
next-generation DNA sequencing platforms.
We developed a new model, the fixed flow cycle model,
to derive the distributions of sequence length
for a given number of flow cycles
under the general conditions 
where the nucleotide incorporation is probabilistic and
may be incomplete, as in some single-molecule sequencing technologies.
Unlike the previous model, the new model
yields the probability distribution
for the sequence length.
Explicit closed form formulas are derived for the mean and variance
of the distribution.

\end{abstract}


\section{Introduction} \label{S:intro}


The  next-generation  sequencing (NGS) technology
has transformed biology and biomedical research by generating huge amount
of sequence data at unprecedented speed.
The emerging DNA sequencing technology also gave rise to
many interesting mathematical and statistical problems.
One of these problems is the statistical distributions
of various quantities associated with sequencing by synthesis (SBS) 
technology.
Besides its practical applications 
in different stages
in the development and use of NGS technology,
such as
instrument development and testing, 
algorithm and software development,
as well as daily monitoring of sequencing outputs,
the problem is an interesting 
self-contained mathematical problem.

Unlike the traditional Sanger capillary electrophoresis method
using dideoxynucleotide chain-termination,
many of the currently available platforms 
and those that are still under development  
utilize SBS technology.
In SBS platforms, 
nucleotides are added to the reactions repeatedly in a pre-determined 
cyclic manner.
In a particular nucleotide cycle, if the added nucleotide is complementary to
the base in the template (the DNA to be sequenced), then the nucleotide
will be potentially synthesized by the enzyme (DNA polymerase)
and a signal can be detected.
Since we know what nucleotide is added in that cycle,
the base in the template can be determined.
If the dispensed nucleotide is not complementary to the base in the
template, or due to the probabilistic nature of the 
enzymatic synthesis reaction
(see subsection~\ref{sss:incomplete} below),
the nucleotide will not be incorporated in this nucleotide cycle.  
Apparently, 
the length of sequence that can be read from the DNA template is not a 
linear function of the number of nucleotide cycles
but rather depends on the sequence context, the nucleotide flow
order, and the reaction conditions that determine the nucleotide incorporation
probabilities.

We have studied the sequence length distributions of SBS previously
\citep{Kong2009,Kong2009b}. There the problem was approached
indirectly. In that model (see subsection~\ref{sss:FSLM} below) 
we fixed the sequence length
and the probability distribution obtained 
is actually the distribution of \emph{flow cycle} for a given sequence length.
It would be more interesting to obtain  directly the distribution
of \emph{sequence length} for a given flow cycle.
In the previous work
we did obtain the sequence length distribution,
but that had to be done by using some approximate normalization.
In this paper we developed a new model, 
\emph{the fixed sequence length model} (\FSLM),
from which we obtain
\emph{directly} the sequence length distribution at a given flow cycle.
This model is more natural than the previous model as it allows for exact
computation of the associated 
probability distribution, while the previous model only yielded approximations.
Furthermore, the new model will be the basis on which
other statistical quantities of SBS platforms 
will be investigated in future studies.

In subsection~\ref{ss:models} the old and new models are to be discussed.
Before that, some basics of SBS technology will first be
introduced.
After this introduction section,
the complete and incomplete nucleotide incorporation cases will be studied 
in section~\ref{S:cni} and section~\ref{S:icni} respectively.
The main results are Eqs.~\eqref{E:sbs_G},
\eqref{E:sbs_avg},
and \eqref{E:sbs_var}
for the general incomplete nucleotide incorporation conditions,
with Eqs.~\eqref{E:pyro_G},
\eqref{E:pyro_avg} 
and \eqref{E:pyro_var} as their special cases 
for complete nucleotide incorporation conditions.

\subsection{SBS technology}
As mentioned earlier, SBS technology adds four nucleotides 
$A$, $C$, $G$, and $T$ repeatedly in a pre-determined order.
To avoid the unnecessary specification of the
detailed names of the four kinds of nucleotides, in the following we will
use $a$, $b$, $c$, and $d$ to represent any permutations of 
the usual nucleotides $A$, $C$, $G$, and $T$,
as we did previously.

We call each of the four cycles where nucleotides are added
a \emph{nucleotide cycle}.
A \emph{flow cycle} is defined as 
the ``quad cycle'' of successive four nucleotides cycles of \{$abcd$\},
with ``$a$'' the first nucleotide in the flow cycle
and ``$d$'' the last nucleotide in the same flow cycle.

SBS reads the template DNA by synthesizing the base that is complementary
to the current base to be read in the template DNA. 
Nucleotide $A$ is complementary to $T$, and $C$ to $G$.
After the complementary nucleotide
is incorporated, certain kind of signal can be
detected, usually through fluorescence dye attached to the
incorporated nucleotide.
The detected signal is registered with the pre-determined nucleotide cycle,
so the nature of the incorporated nucleotide can be determined.
Ideally, if we ignore noises and errors, and if the added nucleotide
is not complementary to the base in the template DNA,
then in that nucleotide cycle no signal will be created.
If the added nucleotide is complementary to the template, however,
we have two possibilities, depending on what kind of NGS technology
we are dealing with.
In the following we discuss two variations of SBS technologies, 
bulk sequencing and single-molecule sequencing,
and the different degree of completeness of nucleotide incorporation
associated with the two technologies.

\subsubsection{Bulk SBS and complete nucleotide incorporation}
For bulk SBS sequencing technologies such as pyrosequencing,
the template DNAs go through a clonal amplification step before
sequencing. What gets sequenced is actually a collection of
identical template molecules.
During the sequencing process, 
the synchronization between identical individual templates
will be lost gradually, leading to signal decay and sequencing errors.
To avoid this dephasing problem, the reaction reagents and synthesis chemistry
are usually tuned to drive the enzymatic incorporation to completion
at each nucleotide cycle.  
Not only will a single base,
if it is complementary to the template base,
be incorporated,
but also a stretch of identical bases (homopolymers)
will be synthesized in the same nucleotide cycle.
For example, if the template DNA has three $G$'s together, then
in one nucleotide cycle of $C$ three nucleotides of $C$ 
will be  incorporated
to match each $G$ in the template.
In ideal conditions for complete nucleotide incorporation, 
the sequence context and nucleotide flow order
uniquely determine the emitted signals.

For example, for sequence \verb+bddabaaad+,
with complete nucleotide incorporation
the signal will be \{0-1-0-2\}-\{1-1-0-0\}-\{3-0-0-1\}, 
while for another sequence
\verb+abbbdabbc+
with the same length, the signal will be \{1-3-0-1\}-\{1-2-1-0\}, if we assume
that the strength of the signal is proportional to the number of
nucleotides incorporated in each nucleotide cycle 
(Table 1, \citet{Kong2009}).
Here we use the four numbers within the braces to 
indicate the signal strength of each nucleotide cycle within a flow cycle.
For example, the number $3$ in the signal for the second sequence 
(the second nucleotide cycle in the first flow cycle)
corresponds to the three consecutive $b$'s in the second sequence.

\subsubsection{Single-molecule
 SBS and incomplete nucleotide incorporation} 
\label{sss:incomplete}

Some NGS platforms that have been commercially available recently
or are under active development utilize 
single-molecule DNA sequencing (SMS) technology.
Unlike the bulk sequencing platforms, 
SMS does not have a clonal amplification step for the target sequences. 
Instead, a single target molecule is used as the template.
This avoids the problems of bias and errors introduced
in the amplification step.
Another advantage of SMS is that the problem of dephasing associated 
with bulk sequencing methods mentioned above does not exist for SMS.
For this reason, 
the reaction kinetics can be controlled to adjust the rate of 
nucleotide incorporation
to the benefit of sequencing accuracy.
For example, slow reaction kinetics can be used deliberately
to limit incorporation to two or three bases per nucleotide cycle
\citep{Harris2008}.
For a homopolymer region \verb+GGG+, 
bulk sequencing technology will try to incorporate three $C$'s
in one nucleotide cycle; 
for SMS, however, zero, one, two, or three $C$'s
can be incorporated in a single synthesis cycle.
This flexibility in 
incorporation rate can be utilized to increase the resolution
of homopolymer region.
Under these conditions, even if the dispensed nucleotide
for a given cycle is complement to the base in the template,
the nucleotide might not be incorporated in the \emph{current}
nucleotide cycle.  The incorporation of the nucleotide
may be delayed to the next cycle, or the next next cycle, and so on.
Which cycle the nucleotide will be incorporated will depend probabilistically
on the reaction conditions.
Thus, the signal will not be deterministically determined by the
sequence context and flow order alone; in addition it will also
depend on the \emph{nucleotide incorporation probability}.
Under incomplete nucleotide incorporation conditions,
the same template sequence will potentially generate many different signals.
This kind of combinatorial explosion 
makes the question of studying the 
sequence length distribution more complicated than under the 
complete nucleotide incorporation conditions.
There are, however,  mathematics tools that can solve the problem
nicely, and the solutions to both complete  incorporation
and incomplete  incorporation are united,
with the former being a special case of the latter.

\subsection{Fixed sequence length model and fixed flow cycle model}
\label{ss:models}

\subsubsection{Fixed sequence length model (\FSLM)}
\label{sss:FSLM}

Two models can be used to investigate various statistic distributions
of SBS.  One model, which was used in our previous work, can be
termed as \emph{fixed sequence length model} (\FSLM).
In this model the length of the template DNA sequences 
is pre-defined as $n$.
Thus, with $4$ possible bases at each position, 
the total number of sequences for a given $n$ is $4^n$.

For \FSLM{} with a given sequence length $n$,
the minimum number of nucleotide flows is $1$, 
for a stretch of $n$ $a$'s.
The maximum number of nucleotide flows 
under complete nucleotide incorporation conditions
is $3n + 1$,
for sequence \verb|dcbadcba...|.
In this case the maximum flow cycle is $f = \lceil (3n+1)/4 \rceil$.
Thus for a fixed sequence length $n$,
the range of  flow cycle $f$ is $[1,   \lceil (3n+1)/4 \rceil ]$.
For incomplete nucleotide incorporation conditions,
the upper limit of the range is determined by the
nucleotide incorporation probabilities.
%
%

Within this model, we have derived the generating function (GF)
for $\R_i(n, f)$,
the probability that a sequence of length $n$ and ending with nucleotide
$i$, $i = a, b, c, d$, is synthesized
in the first $f$ flow cycles,
with the $n$-th nucleotide being synthesized in flow cycle $f$.
With the assumption that the nucleotides in the target sequence are
independent random variables with probabilities $\A, \B, \C$, and $\D$,
and under complete nucleotide incorporation conditions,
the first few values of $\R_i(n, f)$ are shown in 
Table~\ref{T:fixed_n}.
Here and in the following the tables are arranged so that
each row is for a given sequence length $n$, 
and every four columns are for a given flow cycle $f$,
with each of the four columns within a given flow cycle $f$
corresponding to the nucleotide synthesized,
in the order of $a$, $b$, $c$, $d$.
As we have shown, each \emph{row}
in  Table~\ref{T:fixed_n} sums to $1$,
\[
\sum_{f=1}^\infty \sum_{i \in \{a,b,c,d\}} \R_i(n,f) = 1,
\]
so for  \FSLM, $\R_i(n,f)$ is the probability over the flow cycle $f$
for a fixed sequence length $n$.

\begin{table}
\caption{
The first few values of $\R_i(n, f)$
for \emph{fixed sequence length model} (\FSLM)
for complete nucleotide incorporation case.
This model is used previously
but not in this paper.  
The recursive structure of $\R_i(n, f)$
is readily evident. 
Here $\R_d(2,1) = (\A+\B+\C+\D)\D = \D$.
For each \emph{row}, the sum of  $\R_i(n,f)$ is $1$:
 $\sum_{f=1}^\infty \sum_{i \in \{a,b,c,d\}} \R_i(n,f) = 1$.
}
\label{T:fixed_n}
\begin{tabular}{ c|cccc|cccc }
\hline \hline
 & \multicolumn{4}{|c|}{$f=1$} 
 & \multicolumn{4}{|c}{$f=2$} \\
$n$ & $a$ & $b$ & $c$ & $d$ &
   $a$ & $b$ & $c$ & $d$ \\
\hline
$1$ & $\A$ & $\B$ & $\C$ & $\D$ & $0$ & $0$ & $0$ & $0$  \\
$2$ & $\A^2$ & $(\A+\B)\B$ & $(\A+\B+\C)\C$ & $\D$ & 
        $(\B+\C+\D)\A$ & $(\C+\D) \B$   & $\D \C$   & 0 \\
\hline
\end{tabular}
\end{table}

The value of $\R_i(n, f)$ gives the probability that for
a given sequence length $n$ 
, the sequence can be sequenced in $f$ flow cycles
with the last base as $i$. 
The more interesting question,
the answer to which will have more practical uses,
is the distribution of sequence length for a given flow cycle $f$.
To get the answer to this question we have to look at the
\emph{columns} in Table~\ref{T:fixed_n} instead of rows.
The sum of the entries for each \emph{column} (fixed $f$)
in Table~\ref{T:fixed_n}, 
or every four columns within each $f$,
however,
does not add up to $1$, so we have to normalize the entries
to transform them into probabilities.
For \FSLM, only approximate closed form normalization factors can be
found,
though the errors of the approximation are small and
become negligible when $f$ becomes bigger. 

This somewhat unsatisfactory detour motivated us to develop
a model that directly yields the distribution of sequence length
for a fixed number $f$ of flow cycles.  
The model, \emph{fixed flow cycle model} (\FFCM),
is the main focus of this paper.

\subsubsection{Fixed flow cycle model (\FFCM)}

In this model, instead of fixing the sequence length,
we fix the number of flow cycles.
We assume that the sequence is generated by a random process 
that can create sequences with infinite length, only the first part of which
will be sequenced by $f$ flow cycles.
In other words, 
for a fixed number $f$ of flow cycles, 
we assume that the target sequence is always
longer than that can be sequenced by $f$ cycles.
We also assume that the flow cycle $f$ is always a complete cycle (i.e.,  
the number of nucleotide cycles is always a multiple of $4$).
With \FFCM, the size of sample space becomes infinite.
For example, for $f=1$,
the sequence \verb|ba...| can only get the first base  \verb|b|
sequenced,
while the hypothetical sequence \verb|bbbbbb...|
can in theory get infinite read length for the first cycle. 
Hence for $f=1$, 
in the first case the sequence length $n=1$, 
while for the second case the sequence length $n \rightarrow \infty$.
We're interested in the probability distribution of the sequence length $n$
for a given $f$.

\subsection{Simulations}

For \FSLM{}, there are a finite number of sequences 
($4^n$ for sequence length $n$).
Under complete nucleotide incorporation conditions,
the range of flow cycle is also finite (see subsection~\ref{sss:FSLM}),
so the sample space is finite.
The analytical results can be checked by enumeration over all
possible points in this space. Indeed, all results previously obtained
for \FSLM{} have been checked by enumerations.
For \FFCM{}, however, the sample space is infinite, and we cannot
enumerate all possible configurations in its sample space.
To check the analytical results of \FFCM{} developed in this paper,
we developed a simulation program written in C programming language.
The program can be found in
\url{http://graphics.med.yale.edu/sbs/}.
All the results of this paper have been carefully checked
against simulation results.

\subsection{Notation and Definitions}

As mentioned above, we use 
$a$, $b$, $c$, and $d$ to represent any permutations of 
the usual nucleotides $A$, $C$, $G$, and $T$
to avoid the unnecessary specification of the
detailed names of the four kinds of nucleotides.
The probabilities for the four nucleotides in the target sequence
are denoted as 
$p_a$, $p_b$, $p_c$, and $p_d$, with $\sum_{i \in \{a, b, c, d\}} \p_i =1$. 
We assume that the nucleotides in the target
sequence are independent of each other. 
We always use $f$ for the number of flow cycles, and
$n$ as the length of sequenced reads.
To avoid extra symbols, we often use the bases themselves
when their complements should be used, 
if no ambiguity arises.
For example, if the base in the template to be sequenced is $b$,
then it's the complement of $b$ that can be added chemically,
but we will still use $b$ instead of the \emph{complement of $b$}
in the following developments.

The nucleotide incorporation probabilities are denoted as
$\I_{j}^{(i)}$, where $i=a$, $b$, $c$, and $d$ stands 
for the type of nucleotides,
$j=0$, $1$, $2$, $\dots$ denotes the delayed flow cycle number, with the
current flow cycle as $0$. 
The value of $\I_{j}^{(i)}$ is the probability that the next nucleotide 
of type $i$ 
will be incorporated in the $j$-th cycle from the current flow cycle.
For example, when it is complementary to the template,
if the chance for nucleotide $b$ 
to be incorporated in the current cycle, 
the next cycle, and the next-next cycle
is
$1/3$, $1/2$, and $1/6$, respectively, 
then 
$\I_{0}^{(b)} = 1/3$,
$\I_{1}^{(b)} = 1/2$,
$\I_{2}^{(b)} = 1/6$,
and $\I_{j}^{(b)} = 0$ for $j > 2$.
The complete incorporation situation is a special
case in this notation with $\I_{0}^{(i)} = 1$
and $\I_{j}^{(i)} = 0$ for $j > 0$.
By definition $\sum_{j=0}^\infty \I_{j}^{(i)} = 1$.
The GFs of $\p_i \I_{j}^{(i)}$ is denoted by $\g_i(x)$ ($i=a$, $b$, $c$, and $d$):
\begin{equation} \label{E:g}
 \g_i(x) = \p_i \sum_{j=0}^\infty \I_{j}^{(i)} x^j .
\end{equation}
We note that when $x=1$, $\g_i(1) = \p_i$.
Also, for the complete incorporation situation, $\g_i(x) = \p_i$.

To extract coefficients from expansion of GFs,
we use  notation
$[x^n]f(x)$ to denote the coefficient of $x^n$ in the series of
$f(x)$ in powers of $x$.

In Table~\ref{T:symbols} the common notation used throughout the paper
is summarized in one place.  Some of the detailed definitions
will be deferred to the relevant sections in the following.

\begin{table}
\caption{Summary of notation and definitions.}
\label{T:symbols}
\begin{tabular}{cp{5cm}l}
\hline \hline
 & Definition & Comments\\
\hline
$f$          & flow cycle       & $f=1, 2, \dots$ \\
$n$          & sequence length  & $n=0, 1, \dots$  \\
$a$, $b$, $c$, $d$ & any permutation of the 4 nucleotides \\
$\p_i$     & nucleotide probability &  $i=a, b, c, d$\\
$\I_{j}^{(i)}$ & nucleotide incorporation probability of nucleotide type $i$
                 that the next nucleotide will be
		 incorporated in the $j$-th cycle from the current cycle
                &  $i=a, b, c, d$, $j=0, 1, \dots$;
                 $\sum_{j=0}^\infty \I_{j}^{(i)} = 1$\\
$\w_i$ &  conditional factors of nucleotide type $i$, 
                 for complete nucleotide incorporation & 
                 $i=a, b, c, d$\\
$\w_{j}^{(i)}$ &  conditional factors of nucleotide type $i$,
                 for incomplete nucleotide incorporation 
                 & 
                 $i=a, b, c, d$, $j=0, 1, \dots$ \\
$\PP_i(n, f)$  &  probability 
                 that $f$ flow cycles will 
                  synthesize a sequence of length of $n$ and
                  with the last incorporated nucleotide being $i$
		& $i=a, b, c, d$  \\
$\PP(n, f)$    & sum of $\PP_i(n, f)$ & 
		  $\PP(n, f) = \sum_{i \in\{a,b,c,d\}}\PP_i(n, f)$\\
$\R_i(n, f)$  &  probability 
                  that a sequence of length $n$ 
		  and ending with nucleotide
		  $i$ is synthesized
		  in the first $f$ flow cycles,
		  with the $n$-th nucleotide being synthesized 
		  in flow cycle $f$
		  &		 $i=a, b, c, d$ \\
$\R(n, f)$    & sum of $\R_i(n, f)$ &
		  $\R(n, f) = \sum_{i \in\{a,b,c,d\}}\R_i(n, f)$ \\
$\g_i(x)$     & GF of $\p_i \I_{j}^{(i)}$ &
		 $\g_i(x) = \p_i \sum_{j=0}^\infty \I_{j}^{(i)} x^j$  \\
$\h_i(x)$     & GF of $\w_{j}^{(i)}$ &
		  $\h_i(x) = \sum_{j=0}^\infty \w_{j}^{(i)} x^j$ \\
$\F_i(x)$     & GF of $\PP_i(n, f)$ &
		  $\F_i(x, y) = 
		  \sum_{n=0}^\infty \sum_{f=1}^\infty \PP_i(n, f) x^f y^n$ \\
$\F(x)$       & sum of $\F_i(x)$ & $\F(x) = \sum_{i \in\{a,b,c,d\}} \F_i(x)$ \\
$\G_i(x)$     & GF of $\R_i(n, f)$ &
		  $\G_i(x, y) = 
		  \sum_{n=0}^\infty \sum_{f=1}^\infty \R_i(n, f) x^f y^n$ \\
$\G(x)$       & sum of $\G_i(x)$ & $\G(x) = \sum_{i \in\{a,b,c,d\}} \G_i(x)$ \\
$\e_i$        & elementary symmetric functions of $\p_i$ &
		  $i=1, \cdots, 4$, 
		  Eq.~\eqref{E:esf} \\
$\s_i(x)$        & elementary symmetric functions of $\g_i(x) $ &
		  $i=1, \cdots, 4$, Eq.~\eqref{E:esfg} \\
\hline
\end{tabular}
\end{table}

\section{Complete nucleotide incorporation}
\label{S:cni}

In this section the complete nucleotide incorporation conditions
will be studied.
As we showed previously, 
the key to the solution of \FSLM{} is the set of recursive equations
of $\R_i(n, f)$.  
The set of equations cannot be solved analytically, but
transforming them into GFs leads to exact solutions of these GFs.
The recursive structure is evident in the elements in Table~\ref{T:fixed_n}.

For \FFCM{}, let $\PP_i(n, f)$ denote
the probability that $f$ flow cycles will synthesize a sequence
of length of $n$ and
with the last incorporated nucleotide being $i$,
and let $\PP(n, f) = \sum_{i \in \{a,b,c,d\}} \PP_i(n, f)$ 
be the probability that in $f$ flow cycles the sequenced length is $n$.
For this section only, with complete nucleotide incorporation conditions, 
the sequence length $n$ starts from $1$, since in this case 
the probability $\PP(n, f)$ for $n=0$ is zero.  
For the incomplete nucleotide incorporation case discussed later
in section~\ref{S:icni}, however,
the probability to have $n=0$ is nonzero, 
and the case of $n=0$ will be considered there.

With this definition,
it is evident that we have the following 
simple equations
to link the probabilities $\R_i(n, f)$ of \FSLM{} and 
the probabilities $\PP_i(n, f)$ of \FFCM:
\begin{align}
\label{E:pyro_rel}
 \PP_a(n, f) &= 0 ,\\
 \PP_b(n, f) &= \R_b(n, f) \A , \notag\\
 \PP_c(n, f) &= \R_c(n, f) (\A + \B)  , \notag \\
 \PP_d(n, f) &= \R_d(n, f) (\A + \B + \C). \notag
\end{align}
The first part of Eq.~\eqref{E:pyro_rel} is true 
because the first $f$ flow cycles can never synthesize a sequence ending with
$a$ and of length $n$.
Indeed, irrespective of the type of the $(n+1)$-st nucleotide,
it will always be synthesized in the same flow cycle so the length
cannot be $n$.
The second part of Eq.~\eqref{E:pyro_rel} is true 
because if the first $f$ flow cycles are to synthesize a sequence of
length $n$ and ending with $b$, then the first $n$ nucleotides of the
sequence must be synthesized in the first $f$ flow cycles (which happens
with probability $\R_b(n, f)$), and then no more nucleotide
must be synthesized in the $f$-th flow cycle 
(because otherwise the first $f$ flow cycles will synthesize more than
$n$ nucleotides).
The latter in turn implies that the $(n+1)$-st nucleotide
of the sequence must be $a$, which happens with probability $\A$.
The last two parts of Eq.~\eqref{E:pyro_rel} are true for similar
reasons.

The first few values of $\PP_i(n, f)$ are listed in Table~\ref{T:fixed_f}.
Note the factors $\w_i$ in $\PP_i(n, f)$ of Table~\ref{T:fixed_f}
when they're compared with $\R_i(n, f)$ of Table~\ref{T:fixed_n}.
The four conditional factors are 
\begin{equation} \label{E:cond_fac_pyro}
\w_a = 0,
\quad \w_b = \A,
\quad \w_c = \A + \B, 
\quad \text{and}
\quad \w_d = \A + \B + \C .
\end{equation}

\begin{table}
\caption{
The first few values of $\PP_i(n, f)$
for \emph{fixed flow cycle model} (\FFCM)
under complete nucleotide incorporation conditions. 
Note the factors $\w_i$ when compared with the entries 
$\R_i(n, f)$ in Table~\ref{T:fixed_n}. 
These factors are listed in Eq.~\eqref{E:cond_fac_pyro}.
For each $f$, the sum of the four columns of $\PP_i(n,f)$ is $1$:
 $\sum_{n=0}^\infty \sum_{i \in \{a,b,c,d\}} \PP_i(n,f) = 1$.
}
\label{T:fixed_f}
\begin{tabular}{c|cccc|cccc}
\hline \hline
 & \multicolumn{4}{|c|}{$f=1$} 
 & \multicolumn{4}{|c}{$f=2$} 
 \\
 $n$ & $a$ & $b$ & $c$ & $d$ &
       $a$ & $b$ & $c$ & $d$ \\
\hline
$1$ & $\A \w_a$   & $\B \w_b$        & $\C \w_c$           & $\D \w_d$ & $0$ & $0$ & $0$ & $0$ \\
$2$ & $\A^2 \w_a$ & $(\A+\B)\B \w_b$ & $(\A+\B+\C)\C \w_c$ & $\D \w_d $ & 
        $0$ & $(\C+\D) \B \w_b$   & $\D \C \w_c$   & 0   \\
$\cdots$ & $\cdots$ & $\cdots$   & $\cdots$ & $\cdots$ & 
$\cdots$ & $\cdots$ & $\cdots$   & $\cdots$ \\
\hline
\end{tabular}
\end{table}

The equations of probabilities $\PP_i(n, f)$ and $\R_i(n, f)$
in Eq.~\eqref{E:pyro_rel} can then be transformed into
relations between their GFs:
\begin{align}
\label{E:pyro_F}
 \F_a(x, y) &= 0 , \\
 \F_b(x, y) &= \G_b(x, y) \A , \notag \\
 \F_c(x, y) &= \G_c(x, y) (\A + \B)  , \notag \\
 \F_d(x, y) &= \G_d(x, y) (\A + \B + \C) , \notag
\end{align}
where $\F_i(x, y)$ is the bivariate GF of $\PP_i(n, f)$:
\[
\F_i(x, y) = 
  \sum_{n=0}^\infty \sum_{f=1}^\infty \PP_i(n, f) x^f y^n,
 \qquad i = a, b, c, d,  
\]
and $\G_i(x, y)$ is the bivariate GF of $\R_i(n, f)$:
\[
 \G_i (x, y) 
 = \sum_{n=0}^\infty \sum_{f=1}^\infty \R_i(n, f) x^f y^n ,
 \qquad i = a, b, c, d.  
\]
Since $\G_i (x, y)$'s have been solved previously 
(Eqs.~(4)-(7), \citet{Kong2009}),
$\F_i(x, y)$ can be easily obtained from Eq.~\eqref{E:pyro_F}:
\begin{align}
\label{E:pyro_G_abcd}
 \F_a (x, y) &= 0                         , \\
 \F_b (x, y) &= \frac{\A \B x y}{H}  
        \left[ 1 - (\C + \D) (1-x) y 
                           + \C \D (1-x)^2 y^2
        \right]                                  , \notag\\
 \F_c (x, y) &= \frac{(\A + \B) \C x y}{H} 
        \left[ 1 -  \D (1-x) y\right]           , \notag\\
 \F_d (x, y) &= \frac{(\A + \B + \C) \D x y} {H}   \notag     ,         
\end{align}
where
\[
 H = 1 - y + \e_2 (1-x) y^2 - \e_3 (1-x)^2 y^3 + \e_4 (1-x)^3 y^4 .
\]
Here $\e_i$'s are elementary symmetric functions (ESFs)
of nucleotide probabilities $\p_i$:
\begin{align} \label{E:esf}
  \e_1 &= \A + \B + \C + \D = 1               , \\
  \e_2 &= \A \B + \A \C +  \A \D + \B \C 
  + \B \D + \C \D                             ,\notag\\ 
  \e_3 &= \A \B \C +  
  \A \B \D + \A \C \D + \B \C \D              ,\notag\\
  \e_4 &= \A \B \C \D                         .\notag
\end{align}

We're mainly interested in $\PP(n, f)$, the probability
of read length as $n$ for a given flow cycle $f$.
Since $\PP(n, f)$ is the sum of $\PP_i(n, f)$'s,
the GF of $\PP(n, f)$, $\F (x, y)$, is the sum of
$\F_i (x, y)$'s:
\begin{equation} \label{E:pyro_G}
 \F(x, y) = \sum_{i \in \{a, b, c, d\}} \F_i(x, y)
  = \frac{xy}{H} \left[ \e_2 - \e_3(1-x)y + \e_4(1-x)^2 y^2 \right] .
\end{equation}
In comparison, the $\G(x, y)$ of \FSLM{}, 
which is the sum of $\G_i(x, y)$'s,  is given by
\[
\G(x, y) 
= \frac{xy}{H} [ 1 - \e_2(1-x)y + \e_3(1-x)^2 y^2 - \e_4(1-x)^3 y^3] .
\]
Although $\F(x, y)$ and $\G(x, y)$ look similar to each other,
$\G(x, y)$ is the probability generating function (pgf)
for \FSLM{} where sequence length $n$ is fixed,
while $\F(x, y)$ is the pgf of \FFCM{}
where flow cycle $f$ is fixed.
To check that $\PP(n, f)$ generated by GF in Eq.~\eqref{E:pyro_G} 
is a probability distribution over $n$,
we  see that
\[
 \F(x, 1) = \frac{x}{1-x} = x + x^2 + x^3 + \cdots, 
\]
so that $\sum_{n=1}^{\infty} \PP(n, f) = 1$ for any $f$.
On the other hand,
\[
 \G(1, y) = \frac{y}{1-y} = y + y^2 + y^3 + \cdots,
\]
so that $\sum_{f=1}^{\infty} \R(n, f) = 1$ for any $n$.

\begin{figure}
  \centering
  \includegraphics[angle=270,width=\columnwidth]{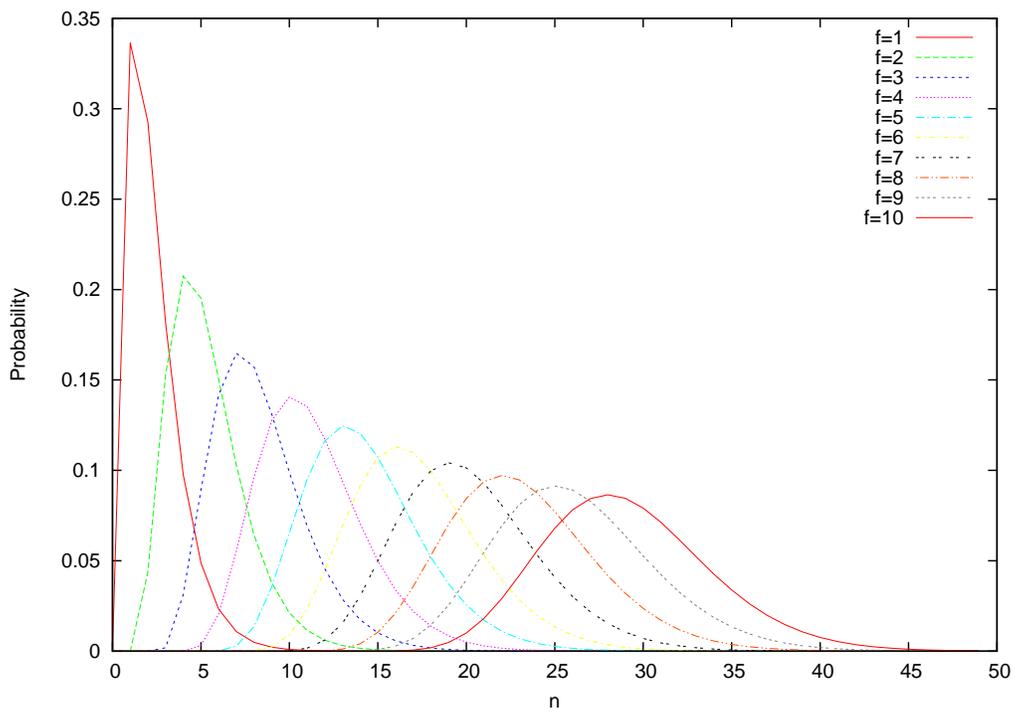}
  \caption{
    The distribution of sequence length for the first $10$ cycles 
    ($f=1,2, \dotsc, 10$)
    with complete nucleotide incorporation.
    The nucleotide composition probabilities used here are
    $\p_a = 1/3 = 0.3333$, $\p_b = 1/11 = 0.0909$, $\p_c = 100/231 = 0.4329$,
    and $\p_d = 1/7 = 0.1429$.
    The distribution is calculated from Eq.~\eqref{E:pyro_G}.
    \label{F:pyro_f1_10}} 
\end{figure}

From Eq.~\eqref{E:pyro_G_abcd} or Eq.~\eqref{E:pyro_G}
we can in principle calculate exactly $\PP_i(n, f)$ or $\PP(n, f)$
for any $f$ and $n$.
In Figure~\ref{F:pyro_f1_10}
the distribution of sequence length are shown for the first $10$ flow cycles 
($f=1,2, \dotsc, 10$).
The nucleotide composition probabilities used here are
$\A = 1/3 = 0.3333$, $\B = 1/11 = 0.0909$, $\C = 100/231 = 0.4329$,
and $\D = 1/7 = 0.1429$.
The distribution is calculated from Eq.~\eqref{E:pyro_G}.


\subsection{Mean and variance}

From GF Eq.~\eqref{E:pyro_G} 
the mean and variance of the sequence length distribution $\PP(n, f)$ 
at a given number of flow cycles $f$ can be calculated:
\begin{subequations}
 \label{E:avg_var_fixed_cycle}
\begin{align}
 \bar{n} (f)  &=  [x^f] \left. \frac{\pd \F(x, y) } {\pd y} \right |_{y=1}  ,  
 \label{E:avg_fixed_cycle}\\
 \var^2 (f) &=  [x^f] \left. \frac{\pd^2 \F(x, y)} {\pd y^2} \right |_{y=1} 
 + \bar{n}(f) - \bar{n}^2(f) .  
\label{E:var_fixed_cycle}
\end{align}
\end{subequations}

To get closed form formulas for $\bar{n} (f)$ and $\var^2 (f)$, we note that 
$x=1$ is a singularity in both 
$ \left. \pd \F(x, y)  / \pd y  \right |_{y=1} $
and
$ \left. \pd^2 \F(x, y)  / \pd y^2  \right |_{y=1} $. 
The denominator of Eq.~\eqref{E:avg_fixed_cycle} has a $(x-1)^2$ factor:
\[
(x-1)^2 (\e_2 - \e_3 + \e_4 + (\e_3 - 2 \e_4) x + \e_4 x^2 ),
\]
and the denominator of Eq.~\eqref{E:var_fixed_cycle}
has a $(x-1)^3$ factor:
\[
(x-1)^3 (\e_2 - \e_3 + \e_4 + (\e_3 - 2 \e_4) x + \e_4 x^2)^2.
\]
For both Eq.~\eqref{E:avg_fixed_cycle} and Eq.~\eqref{E:var_fixed_cycle}
$x=1$ is the pole that has the smallest module.
By using only the principal part of the series expansion, we obtain
the closed form formulas for $\bar{n} (f)$ and $\var^2 (f)$ as
\begin{equation} \label{E:pyro_avg}
 \bar{n} (f)  \approx  \frac{f}{\e_2} + \frac{\e_3}{\e_2^2} - 1,
\end{equation}
and
\begin{equation} \label{E:pyro_var}
  \var^2 (f) \approx \frac{\e_2 + 2\e_3 - 3 \e_2^2}{\e_2^3} f
 - \frac{\e_2^2 \e_3  -5 \e_3^2 + 4 \e_2 \e_4 }{\e_2^4} .
\end{equation}
When these formulas are compared with the corresponding closed form formulas
of \FSLM{} (Eq.~(20) and Eq.~(21), \citet{Kong2009}),
we see that again both the mean and variance
are linear functions of $f$.
Furthermore, the coefficients of $f$ are the same in both models.
The only differences are in the constant terms.

For the special case of equal base probability
where $\A = \B = \C = \D = 1/4$, we have
\begin{align*}
\bar{n} (f) &\approx \frac{8}{3} f - \frac{5}{9}     , \\
\var^2 (f) &\approx \frac{40}{27} f + \frac{20}{81} .
\end{align*}

In Table~\ref{T:pyro_exact_approx}
the differences between the approximate and exact values of 
the mean (Eq.~\eqref{E:pyro_avg}) and the variance 
(Eq.~\eqref{E:pyro_var})
are shown for the first few values of $f$, 
with the same parameters of nucleotide probabilities 
as those in Figure~\ref{F:pyro_f1_10}.
The exact values are calculated from Eq.~\eqref{E:avg_fixed_cycle}
and Eq.~\eqref{E:var_fixed_cycle}.
As we can see, the approximations are very good: for flow cycle as
small as $f=5$ the approximate and exact values are
already very close.

\begin{table}
\caption{
Comparisons of the closed form approximations and exact values 
of the mean (Eq.~\eqref{E:pyro_avg}) and the variance (Eq.~\eqref{E:pyro_var})
for the first few values of $f$, 
with the same parameters as those in Figure~\ref{F:pyro_f1_10}.
    The exact values are calculated from Eq.~\eqref{E:avg_fixed_cycle}
and Eq.~\eqref{E:var_fixed_cycle}.
}
\label{T:pyro_exact_approx}
\begin{tabular}{c|cc|cc}
\hline \hline
& 
\multicolumn{2}{c|}{$\bar{n} (f)$}  & \multicolumn{2}{c}{$\var^2 (f)$} \\
$f$ &   Eq.~\eqref{E:pyro_avg} & exact mean & Eq.~\eqref{E:pyro_var} 
& exact variance\\
\hline
1 &	2.35859351 &	2.39446565 &	2.37546292 &	2.25930624    \\
2 &	5.33118557 &	5.32877823 &	4.58883999 &	4.60388485    \\
3 &	8.30377762 &	8.30387577 &	6.80221706 &	6.80137770    \\
4 &	11.27636968 &	11.27637169 &	9.01559413 &	9.01555228    \\
5 &	14.24896173 &	14.24896084 &	11.22897120 &	11.22898734   \\
6 &	17.22155379 &	17.22155390 &	13.44234827 &	13.44234604   \\
7 &	20.19414585 &	20.19414584 &	15.65572534 &	15.65572555   \\
8 &	23.16673790 &	23.16673790 &	17.86910241 &	17.86910240   \\
9 &	26.13932996 &	26.13932996 &	20.08247948 &	20.08247948   \\
10 &	29.11192201 &	29.11192201 &	22.29585655 &	22.29585655   \\
\hline
\end{tabular}
\end{table}

\begin{figure}
  \centering
  \includegraphics[angle=270,width=\columnwidth]{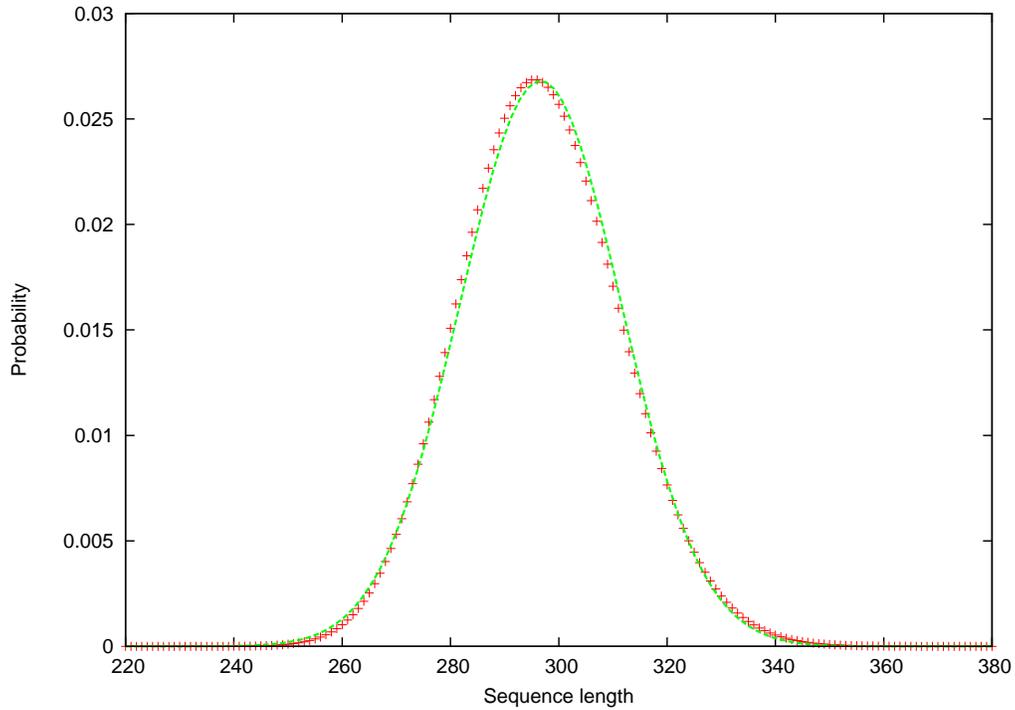}
  \caption{
    The distribution of sequence length with complete nucleotide incorporation
    for a fixed flow cycle of $f=100$.
    The nucleotide composition probabilities used here are the same as
    in Figure~\ref{F:pyro_f1_10}.
    The exact distribution is plotted as '+' and 
    is calculated from Eq.~\eqref{E:pyro_G}.
    The continuous curve is 
    the normal distribution $N({\bar n} (f), \var^2 (f))$
    of the same mean and variance as those of the exact distribution,
    where ${\bar n} (f)$ and $\var^2 (f)$ are calculated
    from Eqs.~\eqref{E:pyro_avg} and \eqref{E:pyro_var}.
    The normal distribution shown here is
    $N(296.6452, 221.4998)$.
    \label{F:pyro_f100}} 
\end{figure}

In Figure~\ref{F:pyro_f100}
the distribution of sequence length $\PP(n, f)$
with complete nucleotide incorporation
for flow cycle $f=100$ is shown,
together with a normal distribution with the same mean and variance.
The nucleotide composition probabilities used here are the same as
in Figure~\ref{F:pyro_f1_10}.
The exact distribution is plotted as '+' and 
is calculated from Eq.~\eqref{E:pyro_G}.
The continuous curve is 
the normal distribution $N({\bar n} (f), \var^2 (f))$
of the same mean and variance as those of the exact distribution
$\PP(n, f)$,
where ${\bar n} (f)$ and $\var^2 (f)$ are calculated
from Eqs.~\eqref{E:pyro_avg} and \eqref{E:pyro_var}.
The normal distribution shown in Figure~\ref{F:pyro_f100} is
$N(296.6452, 221.4998)$.
As Figure~\ref{F:pyro_f100} shows, 
the normal distribution
fits the distribution of $\PP(n, f)$ quite well,
with the exact distribution slightly skews to the left and has
a slightly thick tail on the right.
The approximate normality of the number of nucleotides synthesized
over $f$ flow cycles is not surprising if we make a
central limit theorem type of argument. This number is the sum
of the number of nucleotides synthesized in each of the $f$ flow cycles.
Although these numbers are not independent (because the number of 
nucleotides synthesized in the $(f+1)$-st cycle depends on
the last nucleotide incorporated in the $f$-th cycle), this dependence
is weak and hence the sum will converge to a normal distribution
when $f$ increases.

\section{Incomplete nucleotide incorporation}
\label{S:icni}

For the case of incomplete nucleotide incorporation, where probabilistically
the nucleotides may or may not be incorporated in a given nucleotide flow cycle
based on the enzyme reaction conditions,
the relations become more complicated.
In the following we will express the probability $\PP_i(n, f)$ of \FFCM{}
for the case of incomplete nucleotide incorporation in terms of 
$\R_i(n, f)$, the probability of \FSLM{} 
under incomplete nucleotide incorporation conditions. 
We should emphasize that in the following,
both $\PP_i(n, f)$ and $\R_i(n, f)$ refer to the case
of incomplete nucleotide incorporation. 
The expression of $\R_i(n, f)$ in the following is not the same as that used in
section~\ref{S:cni},
but rather that obtained in \citet{Kong2009b}.
The nucleotide incorporation is now based on the nucleotide incorporation
probabilities $\I_j^{(i)}$, 
and some flow cycles may be skipped for the incorporation.
For $\R_i(n, f)$, however, it's still required that the $n$-th nucleotide
be synthesized in flow cycle $f$, as required by \FSLM{}.

\subsection{Relation between $\PP_i(n, f)$ and $\R_i(n, f)$}

To establish the relation between $\PP_i(n, f)$ and $\R_i(n, f)$, we define
events
\begin{align*}
  A_i(n, f)  &=  \text{\parbox[t]{9.5cm}{
      the first $n$ nucleotides of the sequence are 
      synthesized in the first $f$ flow cycles, with the $n$-th
      nucleotide, of type $i$, being synthesized in flow cycle $f$;}}\\
  B_i(j, f) &=  \text{\parbox[t]{9.5cm}{
      following a synthesis of a nucleotide of type $i$ in flow cycle $f-j$, 
      no more nucleotides are synthesized in flow cycles
      $f-j, f-j+1, \dotsc, f$.}}
\end{align*}
Then, with $\pr$ denoting probability,
$\R_i(n, f) = \pr(A_i(n, f))$, and,
formally by the law of total probability,
\begin{equation} \label{E:conv}
 \PP_i(n, f) = \sum_{j=0}^{f-1} \pr(A_i(n, f-j)) \pr(B_i(j, f))
 = \sum_{j=0}^{f-1} \R_i(n, f-j) \pr(B_i(j, f)) .
\end{equation}
Furthermore the probabilities $\pr(B_i(j, f))$, which we denote by 
$\w_j^{(i)}$, can be expressed as
\begin{align} \label{E:omega}
 \w_{j}^{(a)} & = \A (1-\sum_{k=0}^{j  } \I_{k}^{(a)} ) + \B (1-\sum_{k=0}^{j  } \I_{k}^{(b)} ) + \C (1-\sum_{k=0}^{j  } \I_{k}^{(c)} ) + \D (1-\sum_{k=0}^{j  } \I_{k}^{(d)} ),  \\   
 \w_{j}^{(b)} & = \A (1-\sum_{k=0}^{j-1} \I_{k}^{(a)} ) + \B (1-\sum_{k=0}^{j  } \I_{k}^{(b)} ) + \C (1-\sum_{k=0}^{j  } \I_{k}^{(c)} ) + \D (1-\sum_{k=0}^{j  } \I_{k}^{(d)} ), \notag  \\   
 \w_{j}^{(c)} & = \A (1-\sum_{k=0}^{j-1} \I_{k}^{(a)} ) + \B (1-\sum_{k=0}^{j-1} \I_{k}^{(b)} ) + \C (1-\sum_{k=0}^{j  } \I_{k}^{(c)} ) + \D (1-\sum_{k=0}^{j  } \I_{k}^{(d)} ),  \notag \\   
 \w_{j}^{(d)} & = \A (1-\sum_{k=0}^{j-1} \I_{k}^{(a)} ) + \B (1-\sum_{k=0}^{j-1} \I_{k}^{(b)} ) + \C (1-\sum_{k=0}^{j-1} \I_{k}^{(c)} ) + \D (1-\sum_{k=0}^{j  } \I_{k}^{(d)} )  \notag .
\end{align}
The first part of Eq.~\eqref{E:omega} is true because if the nucleotide
synthesized in flow cycle $f-j$ is of type $a$, 
then the probability that no nucleotide of type $i$ is incorporated 
in flow cycles 
$f-j, f-j+1, \dotsc, f$ is $\p_i (1-\sum_{k=0}^{j  } \I_{k}^{(i)} )$,
 irrespective of the nucleotide type $i$.
The second part of Eq.~\eqref{E:omega} is true because after a nucleotide
of type $b$ is synthesized in flow cycle $f-j$, if the next nucleotide is
of type $a$, it cannot be synthesized in flow cycle $f-j$. 
The probability that it is not incorporated in flow cycles 
$f-j+1, f-j+2, \dotsc, f$ is $\A (1-\sum_{k=0}^{j-1} \I_{k}^{(a)} )$.
If the next nucleotide is of type $i \in \{b, c, d\}$,
then the probability that it is not incorporated in flow cycles 
$f-j, f-j+1, \dotsc, f$ is $\p_i (1-\sum_{k=0}^{j  } \I_{k}^{(i)} )$.
The last two parts of Eq.~\eqref{E:omega} are true for similar reasons.

Eq.~\eqref{E:omega} can be written in a more compact form as
\begin{align} \label{E:omegaII}
 \w_{j}^{(a)} & = 1 - (
   \A \sum_{k=0}^{j  } \I_{k}^{(a)} 
 + \B \sum_{k=0}^{j  } \I_{k}^{(b)}   
 + \C \sum_{k=0}^{j  } \I_{k}^{(c)}  
 + \D \sum_{k=0}^{j  } \I_{k}^{(d)} ) , \\   
 \w_{j}^{(b)} & = \w_{j}^{(a)} + \A \I_{j}^{(a)} ,  \notag \\
 \w_{j}^{(c)} & = \w_{j}^{(b)} + \B \I_{j}^{(b)} ,  \notag \\
 \w_{j}^{(d)} & = \w_{j}^{(c)} + \C \I_{j}^{(c)} .  \notag
\end{align}
For the case of complete incorporation, 
we have $\I_{0}^{(i)} = 1$ and $\I_{j}^{(i)} = 0$, 
$j > 0$, so
$\w_{0}^{(i)}$ reduces to the factors $\w_i$ listed in 
Eq.~\eqref{E:cond_fac_pyro}
while $\w_{j}^{(i)} = 0$ for $j > 0$.

\subsubsection{The special case of $n=0$}
For incomplete incorporation, 
it's possible that for a given flow cycle $f$,
not a single base is synthesized,  
i.e., $n=0$.
This case will never occur in the complete incorporation situation.
By slightly extending the definition of $\w_{j}^{(a)}$ described above,
we can have for $n=0$ and any $f \geqslant 1$,
\begin{align*}
 \PP_a(0, f) &=  \w_{f-1}^{(a)}, \\
 \PP_b(0, f) &=  0, \\
 \PP_c(0, f) &=  0, \\
 \PP_d(0, f) &=  0 .
\end{align*}

\subsection{Solution of $\PP_i(n, f)$}

From Eq.~\eqref{E:conv} we see that
the probabilities of \FFCM{} and \FSLM{} have the following
convolution relation:
\begin{equation} \label{E:convolution}
  \PP_i(n, f) = \sum_{j=0}^{f-1} \R_i(n, f-j) \w_{j}^{(i)}, 
  \qquad i = a, b, c, d.
\end{equation}
Since $\R_i(n, f)$ has been solved \citep{Kong2009b} and $\w_{j}^{(i)}$ is a
function of $\p_i$ and $\I_j^{(i)}$ (Eq.~\eqref{E:omega}),
in principle we can calculate $\PP_i(n, f)$ from Eq.~\eqref{E:convolution}.
To gain insight into $\PP_i(n, f)$, however, we would like to get the
GF of $\PP_i(n, f)$.
The convolution relations in Eq.\eqref{E:convolution} 
transform directly to products of GFs \citep{Wilf2006}. 
Before we transform Eq.\eqref{E:convolution} into GFs,
we first define and calculate the GF of $\w_{j}^{(i)}$.

\subsubsection{Generating function of $\w_{j}^{(i)}$}

If we define the GF of $\w_{j}^{(i)}$, $j=0, 1, 2, \dots$, as
\[ 
  \h_i(x) = \sum_{j=0}^\infty \w_{j}^{(i)} x^j, \qquad i = a, b, c, d,
\]
and using the fact that the GF of
$
 \p_i \sum_{k=0}^{j} \I_{k}^{(i)}   
$
is $\g_i/(1-x)$ \citep{Wilf2006},
then from Eq.~\eqref{E:omega} or Eq.~\eqref{E:omegaII}
we obtain $\h_i(x)$ as
\begin{align} \label{E:h}
  \h_a(x) &= \frac{1 - (\g_a + \g_b + \g_c + \g_d)}{1-x}, \\
  \h_b(x) &= \h_a(x) + \g_a(x) = \frac{1 - (x \g_a + \g_b + \g_c + \g_d)}{1-x}  ,  \notag \\
  \h_c(x) &= \h_b(x) + \g_b(x) = \frac{1 - (x \g_a + x \g_b + \g_c + \g_d)}{1-x}  , \notag \\
  \h_d(x) &= \h_c(x) + \g_c(x) = \frac{1 - (x \g_a + x \g_b + x \g_c + \g_d)}{1-x}  , \notag
\end{align}
where $\g_i(x)$ is the GF of the nucleotide incorporation probabilities
$ \I_{j}^{(i)}$ (times $\p_i$) of nucleotide type $i$,
as defined in Eq.~\eqref{E:g}.
For the complete incorporation case, $\g_i(x) = \p_i$, and $\h_i(x)$'s 
reduce to
\begin{align} \label{E:h0}
 \h_a &= 0, \\
 \h_b &= \p_a, \notag \\
 \h_c &= \p_a + \p_b, \notag \\
 \h_d &= \p_a + \p_b + \p_c , \notag
\end{align}
which agree with the factors $\w_i$ in 
Eq.~\eqref{E:cond_fac_pyro}.

\subsubsection{GF solution of $\PP_i(n, f)$}

Since $\R_i(n, 0) = 0$, 
from the relation between $\PP_i(n, f)$ and $\R_i(n, f)$
of Eq.~\eqref{E:convolution}
we have
\begin{align} \label{E:GFs_gen}
  \F_a (x, y) &= \G_a(x, y)  \h_a(x) + x \h_a(x), \\
  \F_b (x, y) &= \G_b(x, y)  \h_b(x), \notag \\
  \F_c (x, y) &= \G_c(x, y)  \h_c(x), \notag \\
  \F_d (x, y) &= \G_d(x, y)  \h_d(x) .  \notag
\end{align}
The term $x \h_a(x)$ in $\F_a (x, y)$ takes into account of the fact that
$\PP_a(0, f) =  \w_{f-1}^{(a)}$ and we count $f$ from $1$ 
while $\w_{j}^{(i)}$
starts from $j=0$.
We already solved $\G_i(x, y)$ previously as \citep{Kong2009b}
\begin{align*}
 \G_a (x, y) &= \frac{\g_a x y} {H} F    ,        \\
 \G_b (x, y) &= \frac{\g_b x y }{H}  
 \left[ 1 - (\g_c + \g_d) (1-x) y  
   + \g_c \g_d (1-x)^2 y^2  
 \right]                                 ,        \\
 \G_c (x, y) &= \frac{\g_c x y}{H} 
 \left[ 1 -   \g_d (1-x) y \right]       ,        \\
 \G_d (x, y) &= \frac{\g_d x y} {H}      ,        
\end{align*}
where
\[
 H = 1 - \s_1 y + \s_2 (1-x) y^2 - \s_3 (1-x)^2 y^3 + \s_4 (1-x)^3 y^4 ,
\]
and
\[
 F = 
   [ 1 - (\g_b + \g_c + \g_d) (1-x) y  
     + (\g_b \g_c + \g_b \g_d + \g_c \g_d) (1-x)^2 y^2 
     - \g_b \g_c \g_d (1-x)^3 y^3
   ] .
\]
Here $\s_i(x)$'s are elementary symmetric functions (ESFs) of $\g_i(x)$'s,  
the GFs of the nucleotide incorporation probabilities:
\begin{align} \label{E:esfg}
  \s_1 (x) &= \g_a + \g_b + \g_c + \g_d ,                       \\
  \s_2 (x) &= \g_a \g_b + \g_a \g_c +  \g_a \g_d + 
  \g_b \g_c + \g_b \g_d + \g_c \g_d ,\notag\\ 
  \s_3 (x) &= \g_a \g_b \g_c +  \g_a \g_b \g_d + \g_a \g_c \g_d + 
  \g_b \g_c \g_d ,    \notag\\
  \s_4 (x) &= \g_a \g_b \g_c \g_d .                                  \notag
\end{align}
Based on Eq.~\eqref{E:h0}, we see that
 $\F_i(x, y)$ in Eq.~\eqref{E:GFs_gen} reduces to Eq.~\eqref{E:pyro_F}
for the complete incorporation case.

Substitutions of $\G_i (x, y)$ into Eq~\eqref{E:GFs_gen} leads to
the main result we are after:
\begin{equation} \label{E:sbs_G}
 \F(x, y) 
  = \frac{x [1 - \s_1 + \s_2(1-x)y - \s_3(1-x)^2 y^2 + \s_4 (1-x)^3 y^3 ] }
 {(1-x) [ 1 - \s_1 y + \s_2(1-x) y^2 - \s_3 (1-x)^2 y^3 + \s_4 (1-x)^3 y^4 ]} .
\end{equation}
This result is to be compared with the GF of $\R(n,f)$ from \FSLM{}:
\[
 \G(x, y) 
  = \frac{xy [ \s_1 - \s_2(1-x)y + \s_3(1-x)^2 y^2 - \s_4(1-x)^3 y^3]}
  {1 - \s_1 y + \s_2(1-x) y^2 - \s_3 (1-x)^2 y^3 + \s_4 (1-x)^3 y^4}.
\]

To check that $\PP_i(n, f)$ is a probability distribution function over $n$,
we put $y=1$ into Eq.~\eqref{E:sbs_G} to have
\[
  \F(x, 1) = \frac{x}{1-x} = x + x^2 + \cdots,
\]
so the sum over $n$ of values $\PP(n, f)$
for a fixed number $f$ of flow cycles add up to $1$:
$\sum_{n=0}^{\infty} \PP(n, f) = 
 \sum_{n=0}^{\infty} \sum_{i \in \{a,b,c,d\}} \PP_i(n,f)= 1$.
In contrast,
\[
  \G(1, y) = \frac{y}{1-y} = y + y^2 + \cdots.
\]

\begin{figure}
  \centering
  \includegraphics[angle=270,width=\columnwidth]{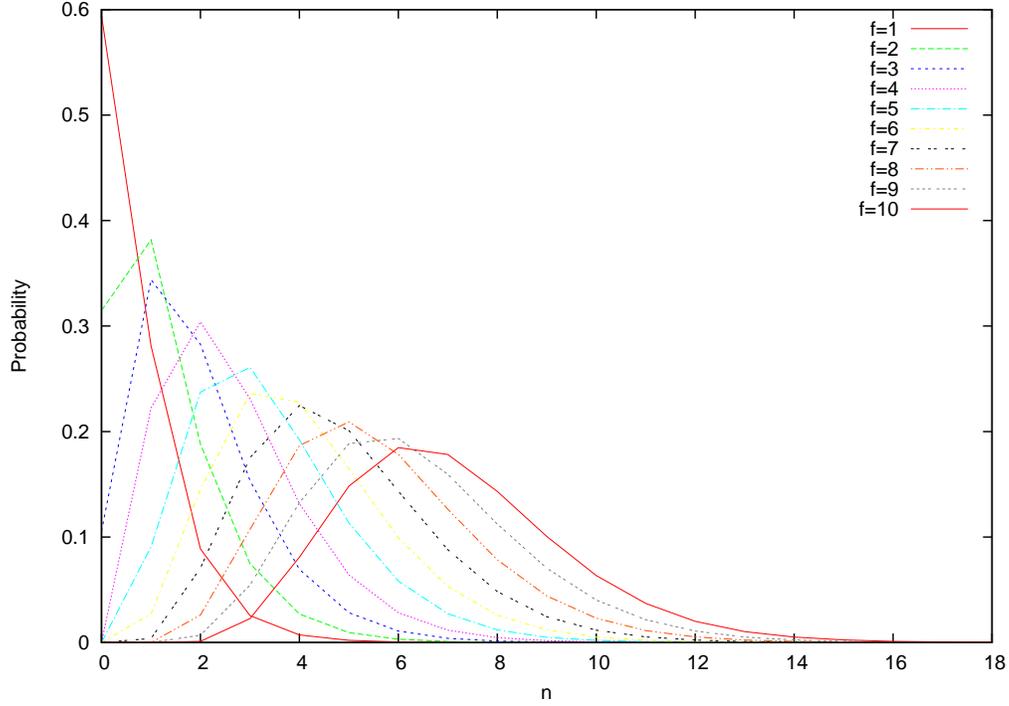}
  \caption{
    The distribution of sequence length for the first $10$ cycles
    ($f=1,2, \dots, 10$)
    with incomplete nucleotide incorporation.
    The nucleotide composition probabilities used here are
    the same as in Figure~\ref{F:pyro_f1_10}.
    The hypothetical non-zero nucleotide incorporation probabilities are
    $\I_{{j}}^{(a)} = [6/55, 1/2, 3/10, 1/11]$,
    $\I_{j}^{(b)} = [19/60, 1/4, 1/3, 1/10]$,
    $\I_{j}^{(c)} = [407/630, 1/7, 1/10, 1/9]$,
    and 
    $\I_{j}^{(d)} = [17/40, 1/5, 1/4, 1/8]$.
    The distribution is calculated from Eq.~\eqref{E:sbs_G}.
    \label{F:sbs_f1_10}} 
 \end{figure}

The distributions of sequence length for the first $10$ flow cycles
($f=1,2, \dots, 10$)
with incomplete nucleotide incorporation 
are show in Figure~\ref{F:sbs_f1_10}.
 The nucleotide composition probabilities used are
 $\p_a = 1/3 = 0.3333$, $\p_b = 1/11 = 0.0909$, $\p_c = 100/231 = 0.4329$,
 and $\p_d = 1/7 = 0.1429$,
 the same as those used for complete nucleotide incorporation case,
 such as Figure~\ref{F:pyro_f1_10}.
 The hypothetical non-zero nucleotide incorporation probabilities used are
 $\I_{{j}}^{(a)} = [6/55, 1/2, 3/10, 1/11]$,
 $\I_{j}^{(b)} = [19/60, 1/4, 1/3, 1/10]$,
 $\I_{j}^{(c)} = [407/630, 1/7, 1/10, 1/9]$,
 and 
 $\I_{j}^{(d)} = [17/40, 1/5, 1/4, 1/8]$,
 for $j=0$, $1$, $2$, and $3$. 
 For $j>3$, $\I_{j}^{(i)} = 0$.
The distribution is calculated from Eq.~\eqref{E:sbs_G}.
These distributions can be compared with those
in Figure~\ref{F:pyro_f1_10} of complete nucleotide incorporation conditions.

\subsection{Mean and variance}
From the GF $\F(x,y)$ we can obtain closed form 
expressions of the mean and variance of the sequence length distribution using 
Eq.~\eqref{E:avg_fixed_cycle} and Eq.~\eqref{E:var_fixed_cycle}.
Since $\s_1(1) = 1$, 
the denominator of $ \left. \pd \F(x, y)  / \pd y  \right |_{y=1} $ has a $(x-1)^2$ factor,
and 
the denominator of $ \left. \pd^2 \F(x, y)  / \pd y^2  \right |_{y=1} $ has a $(x-1)^3$ factor.
Again $x=1$ is the pole that has the smallest module.
Using the principal part of the series expansion,
we get the closed form expression of the mean and variance
of $\PP(n, f)$ as
\begin{equation} \label{E:sbs_avg}
 \bar{n} (f)  \approx  
   \frac{f}{u} - \frac{2 \e_2^2 - 2 \e_3 + 4 \e_2 \s_1' + 2 (\s_1')^2 - \s_1'' - 2 \s_2' }{2 u^2} ,
\end{equation}
and
\begin{equation} \label{E:sbs_var}
 \var^2 (f) \approx 
   \frac{v - (3 \e_2 + \s_1' - 1) u}{u^3} f + \frac{w}{12 u^4},
\end{equation}
where 
\begin{align*}
 u &= \e_2 + \s_1' , \\
 v &= 2 \e_3 + \s_1'' + 2 \s_2' ,\\
 w &= 6 u v (3 u-4 \e_2) + 15 v^2 
      - 8 u [ 3u(\s_2' + \s_1'') + 6 \s_3' + \s_1^{(3)} + 3\s_2'' + 6 \e_4 ] .
\end{align*}
Here we use the abbreviations 
$\s_i'$ to denote $\s_i'(1)$, 
$\s_i''$ as 
$\s_i''(1) = \pd^2 \s_i(x) /\pd x^2 |_{x=1}$, etc.

From Eq.~\eqref{E:sbs_avg} and Eq.~\eqref{E:sbs_var} we see that
both mean and variance are linear function of $f$.
The coefficients of $f$ are the same for both \FFCM{}
and \FSLM{} (Eq.~(17a) and Eq.~(17b) of Ref.~\citet{Kong2009b}).
Only the constant terms differ between the two models.

In Table~\ref{T:sbs_exact_approx}  
comparisons of the closed form approximations of
the mean (Eq.~\eqref{E:sbs_avg}) and the variance (Eq.~\eqref{E:sbs_var})
against their respective exact values for the first few values of
$f$ are shown, 
with the same parameters as those in Figure~\ref{F:sbs_f1_10}.
The exact values are calculated from Eq.~\eqref{E:sbs_G}
using Eq.~\eqref{E:avg_fixed_cycle} and Eq.~\eqref{E:var_fixed_cycle}.
We can see that the closed form approximations are quite close
to the exact values. As $f$ increases, the errors become
smaller; when $f$ reaches $f=8$, the errors become negligible.

\begin{table}
\caption{
  Comparisons of the closed form approximations of
  the mean (Eq.~\eqref{E:sbs_avg}) and the variance (Eq.~\eqref{E:sbs_var})
  against their respective exact values for the first few values of
  $f$, 
  with the same parameters as those in Figure~\ref{F:sbs_f1_10}.
  The exact values are calculated from Eq.~\eqref{E:sbs_G}
  using Eq.~\eqref{E:avg_fixed_cycle} and Eq.~\eqref{E:var_fixed_cycle}.
}
\label{T:sbs_exact_approx}
\begin{tabular}{c|cc|cc}
\hline \hline
& 
\multicolumn{2}{c|}{$\bar{n} (f)$}  & \multicolumn{2}{c}{$\var^2 (f)$} \\
$f$ &   Eq.~\eqref{E:sbs_avg} & exact mean & Eq.~\eqref{E:sbs_var} 
& exact variance\\
\hline
1	&	0.43744094	&	0.57921752	&	0.82083681	&	0.72850788 \\
2	&	1.17769743	&	1.16602712	&	1.29206252	&	1.30509385 \\
3	&	1.91795391	&	1.89883904	&	1.76328823	&	1.80247850 \\
4	&	2.65821039	&	2.68250622	&	2.23451393	&	2.15175889 \\
5	&	3.39846688	&	3.39672794	&	2.70573964	&	2.71085442 \\
6	&	4.13872336	&	4.13533479	&	3.17696535	&	3.19335924 \\
7	&	4.87897985	&	4.88027760	&	3.64819106	&	3.64210312 \\
8	&	5.61923633	&	5.61927551	&	4.11941677	&	4.11866116 \\
9	&	6.35949281	&	6.35913945	&	4.59064248	&	4.59319956 \\
10	&	7.09974930	&	7.09989881	&	5.06186819	&	5.06080288 \\
11	&	7.84000578	&	7.84002311	&	5.53309389	&	5.53286474 \\
12	&	8.58026226	&	8.58022040	&	6.00431960	&	6.00472717 \\
13	&	9.32051875	&	9.32053270	&	6.47554531	&	6.47541913 \\
14	&	10.06077523	&	10.06077853	&	6.94677102	&	6.94672379 \\
15	&	10.80103172	&	10.80102741	&	7.41799673	&	7.41804782 \\
16	&	11.54128820	&	11.54128938	&	7.88922244	&	7.88920999 \\
17	&	12.28154468	&	12.28154513	&	8.36044814	&	8.36044095 \\
18	&	13.02180117	&	13.02180070	&	8.83167385	&	8.83168045 \\
19	&	13.76205765	&	13.76205775	&	9.30289956	&	9.30289844 \\
20	&	14.50231414	&	14.50231420	&	9.77412527	&	9.77412416 \\
\hline
\end{tabular}
\end{table}

\begin{figure}
  \centering
  \includegraphics[angle=270,width=\columnwidth]{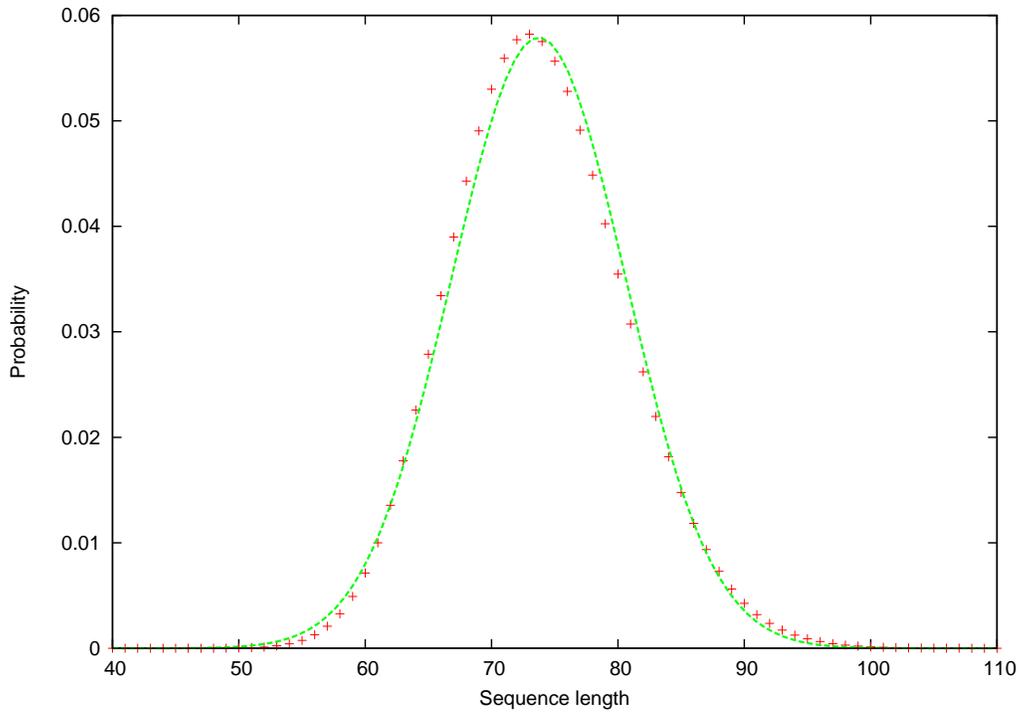}
  \caption{
    The distribution of sequence length with 
    incomplete nucleotide incorporation
    for a fixed flow cycle of $f=100$.
    The nucleotide composition probabilities 
    and the nucleotide incorporation probabilities 
    used here are the same as
    in Figure~\ref{F:sbs_f1_10}.
    The exact distribution is plotted as '+' and 
    is calculated from Eq.~\eqref{E:sbs_G}.
    The continuous curve is 
    the normal distribution $N({\bar n} (f), \var^2 (f))$
    of the same mean and variance as those of the exact distribution,
    where ${\bar n} (f)$ and $\var^2 (f)$ are calculated
    from Eqs.~\eqref{E:sbs_avg} and \eqref{E:sbs_var}.
    The normal distribution shown here is
    $N(73.7228, 47.4722)$.
    \label{F:sbs_f100}} 
\end{figure}

Figure~\ref{F:sbs_f100} shows 
the distribution of sequence length with incomplete nucleotide incorporation
for a fixed flow cycle of $f=100$.
The parameters used here are the same as
in Figure~\ref{F:sbs_f1_10}.
The exact distribution is plotted as '+' and 
is calculated from Eq.~\eqref{E:sbs_G}.
The continuous curve is 
the normal distribution $N({\bar n} (f), \var^2 (f))$
of the same mean and variance as those of the exact distribution,
where ${\bar n} (f)$ and $\var^2 (f)$ are calculated
from Eqs.~\eqref{E:sbs_avg} and \eqref{E:sbs_var}.
The normal distribution shown here is
$N(73.7228, 47.4722)$.
The agreement of the exact distribution with normal distribution is decent,
with a slight skew to the left.

\begin{figure}
  \centering
  \includegraphics[angle=270,width=\columnwidth]{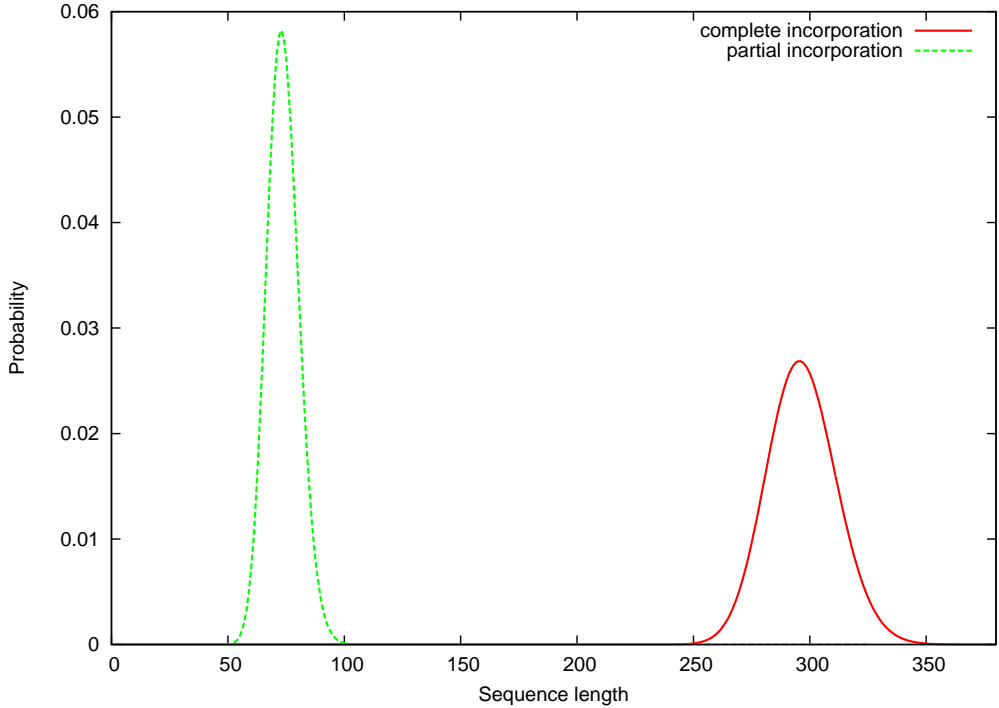}
  \caption{
    The distributions of sequence length
    for a fixed flow cycle  of $f=100$ for
    complete and incomplete nucleotide incorporation.
    The curve on the right is for
    complete nucleotide incorporation,
    the curve on the left is for
    incomplete nucleotide incorporation.
    The nucleotide composition probabilities
    and the nucleotide incorporation probabilities are 
    the same as in Figure~\ref{F:pyro_f100} 
    and Figure~\ref{F:sbs_f100}.
    \label{F:sbs_f100_pyro_and_gen}} 
\end{figure}

For comparison,
the distributions of sequence length with complete and incomplete nucleotide incorporation
for a fixed flow cycle of $f=100$ are shown in the same plot in
Figure~\ref{F:sbs_f100_pyro_and_gen}.
The nucleotide composition probabilities
and the nucleotide incorporation probabilities are the same as those 
in Figure~\ref{F:pyro_f100} and Figure~\ref{F:sbs_f100}.
The delays of synthesis in the  
incomplete incorporation case shift the mean of the sequence length
distribution to the left, with a smaller variance.

\section*{Acknowledgment}
  This work was supported by
  Yale School of Medicine.


\bibliographystyle{spbasic}      


\end{document}